\documentclass[aps,prd,twocolumn,superscriptaddress,amssymb,amsmath,nofootinbib]{revtex4-2}

\usepackage{graphicx}
\usepackage[colorlinks=true,linkcolor=blue,citecolor=blue,urlcolor=blue]{hyperref}

\begin{document}

\title{$Zb\bar b$ couplings at the Large Hadron Collider}

\author{Radovan Dermisek}
\email[]{dermisek@iu.edu}
\affiliation{Department of Physics, Indiana University, Bloomington, Indiana, 47405, USA}

\author{Keith Hermanek}
\email[]{khermane@iu.edu}
\affiliation{Department of Physics, Indiana University, Bloomington, Indiana, 47405, USA}

\author{Marcos A. Hernandez Ojeda}
\email[]{hernma@iu.edu}
\affiliation{Department of Physics, Indiana University, Bloomington, Indiana, 47405, USA}

\author{Taegyu~Lee}
\email[]{taeglee@iu.edu}
\affiliation{Department of Physics, Indiana University, Bloomington, Indiana, 47405, USA}

\author{Navin McGinnis}
 \email[]{nmcginnis@uchicago.edu}
  \affiliation{High Energy Physics Division, Argonne National Laboratory, Lemont, IL 60439, USA}
 \affiliation{Enrico Fermi Institute, Physics Department, University of Chicago, Chicago, IL 60637, USA}
 \affiliation{Department of Physics, University of Arizona, Tucson, Arizona 85721, USA}

\date{November 19, 2026}

\begin{abstract}
Constraining the $Z$-boson couplings of the $b$ quark, or even ruling out the possibility of an opposite-sign coupling, is considered very challenging at the LHC. Utilizing signatures of new physics resulting from contact operators accompanying the modification of $Zb \bar{b}$ couplings from Standard Model predictions, we show that the ATLAS measurement of $pp\to Zh$ already ruled out the opposite-sign couplings by more than 9.5 standard deviations. We also argue that $pp\to Zhb$, which receives a contribution from the same contact term, would lead to even better sensitivity. Furthermore, we show that with future data this indirect method can provide the strongest constraints on $Zb\bar b$ couplings, and can even probe the anomaly in these couplings left over from the LEP era. 
\end{abstract}

\maketitle

\section{Introduction}

Precision tests of the Standard Model of particle physics (SM) include measurements of the couplings of quarks and leptons to gauge bosons, which are all predicted in terms of their quantum numbers and three gauge couplings. Among the least constrained are the couplings of the $b$ quark to the $Z$ boson, measured at LEP and SLC, which indicate $\gtrsim$ $2\sigma$ deviation from the SM prediction (mostly from the right-handed coupling)~\cite{ParticleDataGroup:2026aaa}. Furthermore, the sign of the right-handed coupling has not yet been constrained, leaving two possible solutions in global fits of SM couplings to data~\cite{Choudhury:2001hs,Batell:2012ca} for more than two decades. 

Constraining the $Z$-boson couplings of the $b$ quark, or at least ruling out the possibility of an opposite-sign coupling, has been considered even more challenging at the Large Hadron Collider (LHC) due to substantial QCD background. Limited studies indicate that with the full data set of the high luminosity run (HL-LHC) we might be able just to rule out the opposite-sign coupling~\cite{Beccaria:2012xw, Gauld:2015qha, Gauld:2019doc, Yan:2021veo, Dong:2022ayy, Bishara:2023qhe}. Even that is highly non-trivial, since in order to have access to the sign of the coupling, the interference with a process where the coupling does not enter has to be measured. This is also the reason why the $Z$-pole measurements at LEP could not constrain the signs, and limited measurements off the $Z$-pole were able to determine only the sign of the much larger left-handed coupling.
Thus, most of the focus related to the puzzle in $Zb\bar b$ couplings has been on possible future colliders, for example the Future Circular Collider (FCC) or a linear collider, which could repeat LEP measurements on and off the $Z$-pole with much better precision~\cite{Gori:2015nqa,Bilokin:2017lco,Yan:2023ccj}.

\begin{figure}[t!]
\includegraphics[width=3.2in]{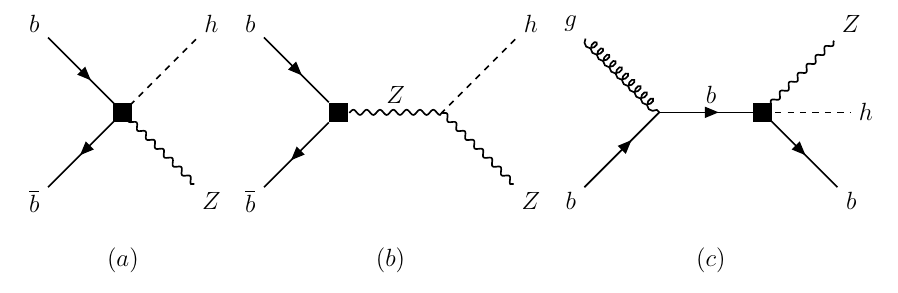}
\caption{Examples of diagrams contributing to $pp \to Zh$  and $pp \to Zhb$  from new physics modifying $Zb\bar b$ couplings.}
\label{fig:smeft_diags}
\end{figure}

In this paper we show that,  utilizing signatures of new physics resulting from contact operators accompanying the modification of $Zb\bar{b}$ couplings from SM predictions, the ATLAS measurement of $pp\to Zh$~\cite{ATLAS:2020fcp} already rules out the opposite-sign couplings at more than 9.5 standard deviations. Focusing on the contribution from the contact $Z$-$h$-$b$-$\bar b$ interactions to this process, Fig.~\ref{fig:smeft_diags} (a), which accompany modifications of $Zb\bar b$ couplings, removes the necessity of measuring the interference with another contribution to the same process. The coupling with the SM sign requires a small or zero Wilson coefficient of the corresponding effective operators, while the opposite-sign coupling requires a substantially large Wilson coefficient enough to produce a correction twice the size of dimension-4 terms in the SM. As such, the opposite-sign scenario is expected to be associated with large contributions to processes where the contact interaction enters, $b\bar b\to Zh$ being the simplest one. We also argue that $pp \to Zhb $, which receives a contribution from the same contact term (see Fig.~\ref{fig:smeft_diags} (c))  would lead to even better sensitivity due to smaller SM background.  Furthermore, we show that with future data this indirect method can provide the strongest constraints on $Zb\bar b$ couplings at the LHC, and can even probe the anomaly in these couplings as a remnant from the LEP era.

The best prospects for constraining the $Zb\bar b$ couplings at the LHC have been previously discussed in Refs.~\cite{Gauld:2019doc, Yan:2021veo, Bishara:2023qhe}. A constraint can be obtained at LHCb through the ratio of the bottom to charm asymmetry in the difference of rapidities of the heavy quark and anti-quark, which takes advantage of LHCb's excellent charge-tagging capabilities~\cite{Gauld:2019doc}. In Ref.~\cite{Yan:2021veo} the authors claim that by accessing the axial-vector $Zb\bar b$ coupling through measurements of the loop induced $gg \to Zh$ (see Fig.~\ref{fig:ggzh_diags} (b)), the opposite-sign solution can be ruled out at HL-LHC and under certain assumptions, comparable bounds to LEP could be achieved. However, we find that the new physics contribution from Fig.~\ref{fig:ggzh_diags} (b) is canceled by the contribution from the contact interaction, Fig.~\ref{fig:ggzh_diags} (a), invalidating this possibility. Such a cancelation was discussed previously for modified top-quark couplings~\cite{Rossia:2023hen}. Finally, in Ref.~\cite{Bishara:2023qhe} it was suggested to measure the charge asymmetry of $b$-jets in $pp\to Zb \bar b$. With current flavor and charge-tagging capabilities, it was claimed that at HL-LHC this method could reject the opposite-sign solution at 4$\sigma$. We will see that focusing on signals from the contact interactions will provide much stronger results.

While our main results assume that only the dimension-6 operators responsible for modifications of $Zb\bar b$ couplings are present,  we discuss that the results remain quite robust even when contributions from other dimension-6 operators are considered. In addition, we comment on the possible impact of operators of dimension-8 or higher and how to constrain them by other searches.

\section{Modified $Zb\bar b$ couplings and $pp\to Zh$,  $Zhb$ signals} 
\label{sec:ModCouplings}

The part of the Lagrangian describing $Zb\bar b$ couplings is given by
\begin{equation}
\mathcal{L} \supset    \frac{g}{\cos \theta_W} \left(  g_L^{Z bb} \ \overline{b}_L \gamma^{\mu} b_L  \, + \, g_R^{Z bb} \  \overline{b}_R \gamma^{\mu} b_R  \right)  Z_{\mu} ,
\label{eq:lagrangian}
\end{equation}
where $g$ is the $SU(2)$ gauge coupling, $ \theta_W$ is the weak mixing angle, and $g_{L,R}^{Z bb} $ can be written as the sum of the SM predictions and contributions of new physics:
\begin{eqnarray}
g_L^{Z bb}  &=&  g_{L, \textrm{SM}}^{Z bb} + \delta g_L^{Z bb}  , \label{eq:gL}\\
g_R^{Z bb}  &=&  g_{R, \textrm{SM}}^{Z bb} + \delta g_R^{Z bb} .
\label{eq:gR}
\end{eqnarray}
The SM values of the couplings are obtained from tree level predictions,
\begin{eqnarray}
g_{L, \textrm{SM}}^{Z bb (\textrm{tree})}  &=&  - \frac{1}{2} + \frac{1}{3} \sin^2 \theta_W  , \\
g_{R, \textrm{SM}}^{Z bb (\textrm{tree})}  &=& \frac{1}{3} \sin^2 \theta_W ,
\label{eq:g-tree}
\end{eqnarray}
which after radiative corrections result in $g_{L,\textrm{SM}}^{Z bb} = -0.4196$ and $g_{R, \textrm{SM}}^{Z bb} =0.0771 $~\cite{ParticleDataGroup:2026aaa}. The contributions of new physics parametrized by operators in the SM effective field theory (SMEFT) can be written as 
\begin{eqnarray}
\delta g_L^{Z bb}  &=&  -\frac{v^2}{2 \Lambda^2} \left( C_{H q}^{(1) 33} + C_{Hq}^{(3) 33} \right) + {\mathcal{O}}\left( \frac{v^4}{ \Lambda^4}\right), 
\label{eq:del_gL} \\
\delta g_R^{Z bb}  &=& -\frac{v^2}{2 \Lambda^2} \ C_{Hd}^{33} + {\mathcal{O}}\left( \frac{v^4}{ \Lambda^4}\right),
\label{eq:del_gR}
\end{eqnarray}
where the $v^2 /  \Lambda^2$-suppressed terms originate from dimension-6 operators $( C_{H q}^{(1) 33 } / \Lambda^2) ( H^{\dagger} i \overleftrightarrow{D}_{\mu} H ) (\overline{q}_L \gamma^{\mu} q_L)  $, $(C_{H q}^{(3) 33} / \Lambda^2 )(H^{\dagger} i \overleftrightarrow{D}^a_{\mu} H) ( \overline{q}_L \tau^a \gamma^{\mu} q_L )$,  and $  ( C_{H d}^{33} / \Lambda^2 ) (H^{\dagger} i \overleftrightarrow{D}_{\mu} H ) ( \overline{b}_R \gamma^{\mu} b_R )$ after the Higgs doublet  develops a vacuum expectation value, $H = (0, (v + h)/\sqrt{2})^T$ with $v = 246$ GeV and $h$ the SM Higgs boson. The quark doublet is $q_L = (t_L, b_L)^T$. Although we follow the standard notation for SMEFT operators in the Warsaw basis~\cite{Grzadkowski:2010es}, we explicitly factor out the cutoff scale $\Lambda$ at which new physics enters significantly above the electroweak scale. Terms originating from operators of dimension-8 and higher at ${\mathcal{O}}(v^4 /  \Lambda^4)$ are neglected in our main results, but we discuss their potential impact in Sec.~\ref{sec:discussion}.

The same dimension-6 operators also generate $Z$-$h$-$b$-$\bar b$ contact interactions, 
\begin{equation}
\mathcal{L} \supset    \frac{2g}{v \cos \theta_W} \left(  \delta g_L^{Z bb} \ \overline{b}_L \gamma^{\mu} b_L  \, + \, \delta g_R^{Z bb} \  \overline{b}_R \gamma^{\mu} b_R  \right)  Z_{ \mu} \, h ,
\label{eq:contact_term}
\end{equation}
proportional to modifications of the $Zb\bar b$ couplings, which provide new opportunities to test them. The simplest example is the $b\bar b\to ZH$
subprocess of $pp\to ZH$, to which the contact terms directly contribute together with the SM diagrams with modified $Zb\bar b$ couplings, see Fig.~\ref{fig:smeft_diags} (a) and (b). In the high-energy limit, the new physics contribution to $pp\to Zh$ is dominated by the contact interactions with the longitudinal mode of the $Z$ boson. Considering only the contribution from the contact interactions, the parton level cross section is given by 
\begin{equation}
\hat{\sigma}_{b\bar b \to Z h}^{\textrm{(NP)}} = \frac{1}{24 \pi v^4} \, F(\hat{s}) \left(|\delta g_L^{Zbb}|^2 + |\delta g_R^{Zbb}|^2\right),
\end{equation}
where $F(\hat{s}) \equiv  (\hat{k}/\sqrt{\hat{s}} ) ( \hat{s} - 4m_Z^2 - 4 \hat{k}^2 / 3)$ with $\hat{k}^2 = ((\hat{s} - (m_h^2 + m_Z^2))^2 - 4 m_Z^2m_h^2)/4 \hat{s}$, and $\hat{s}$ is the parton center-of-mass energy squared. In the high energy limit, $F(\hat{s}) \simeq  \hat{s}/3$ and thus $\hat{\sigma}_{b\bar b \to Z h}^{\textrm{(NP)}}$ grows with energy squared.  On the other hand, the parton level cross section $q \bar q \to Z h$ in the SM is suppressed by energy squared~\cite{Ellis:1975ap}. As a result of the energy growth of the contact term-squared contribution, the interference with the SM contribution is small for sufficiently large Wilson coefficients. Neglecting the interference with the SM contribution, the signal is proportional to $|\delta g_L^{Zbb}|^2 + |\delta g_R^{Zbb}|^2$, and we will see that this is a good approximation to the full results for currently available data at the LHC.

The same contact interactions contribute also to $gb\to Zhb$ (see Fig.~\ref{fig:smeft_diags} (c)) and thus to $pp\to Zhb$, which provide an additional test of the modified $Zb\bar b$ couplings. The energy growth associated with the contact interactions also applies to this process. Thus, the main distinguishing feature of the new physics signal in both $pp\to Zh$ and $pp\to Zhb$ from the corresponding SM background is a preference for $Z$, $h$, and potentially an extra $b$-jet to emerge with larger transverse momenta.

\begin{figure}[t!]
\includegraphics[width=3.2in]{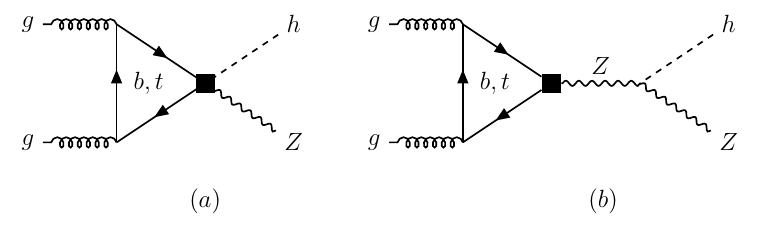}
\includegraphics[scale=0.65]{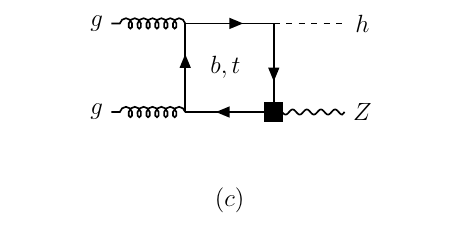}
\caption{Diagrams contributing to $gg \to Zh$ at one loop from new physics modifying $Zb\bar b$ couplings. The operators modifying $g_L^{Z bb}$ also modify $g_L^{Z tt}$, and thus top-loops also contribute.}
\label{fig:ggzh_diags}
\end{figure}

Modified  $Zb\bar b$ couplings also contribute to the $gg \to Zh$ at one loop through diagrams in Fig.~\ref{fig:ggzh_diags}. Since the operators modifying $g_L^{Z bb}$ also modify the $g_L^{Z tt}$,
\begin{equation}
\delta g_L^{Z tt}  =  -\frac{v^2}{2 \Lambda^2} \left( C_{H q}^{(1) 33} - C_{Hq}^{(3) 33} \right) + {\mathcal{O}}\left( \frac{v^4}{ \Lambda^4}\right), 
\label{eq:Ztt}
\end{equation}
the top-loops also contribute. The contribution of the diagram in Fig.~\ref{fig:ggzh_diags} (b) was studied in Ref.~\cite{Yan:2021veo} as a possible way to gain access to the axial-vector $Zb\bar b$ coupling. The authors claimed that at the HL-LHC the opposite-sign solution can be ruled out and, under certain assumptions, comparable bounds to LEP could be achieved. However, we find that the new physics contribution from Fig.~\ref{fig:ggzh_diags} (b) is canceled by the contribution from the contact interaction, Fig.~\ref{fig:ggzh_diags} (a), which was not considered in Ref.~\cite{Yan:2021veo}. Such a cancelation was discussed previously for the modified top-quark couplings~\cite{Rossia:2023hen}. We find that exact cancellation only occurs when the modifications of $Zb\bar b$ couplings determine the $Z$-$h$-$b$-$\bar b$ contact interactions as in Eq.~(\ref{eq:contact_term}), and is thus limited to contributions from operators of dimension-6. Since the proposal in Ref.~\cite{Yan:2021veo}  is not sensitive to the leading contribution of new physics, it cannot be used as an indirect measurement of the axial-vector  $Zb\bar b$ coupling. However, it could be potentially used to constrain contributions of dimension-8 operators and higher, as we discuss in Sec.~\ref{sec:discussion}. 

The bottom quark loop in the remaining contribution to the $gg \to Zh$, Fig.~\ref{fig:ggzh_diags} (c), is negligible due to the small Yukawa coupling. The amplitude for the top loop diagram in Fig.~\ref{fig:ggzh_diags} (c) grows at most as $\sim \delta g_L^{Ztt} \times (m_t^2 / v^2) \ \textrm{ln}^2(\hat{s} / m_t^2)$, and the corresponding contribution to the cross section is suppressed by $\hat s$ from the phase space~\cite{Rossia:2023hen}.

\section{Limits on $Zb\bar b$ couplings from Atlas analysis of $pp\to Zh$}
\label{sec:ATLAS_CURRENT}

We derive limits on $Zb\bar b$ couplings and the corresponding Wilson coefficients from the ATLAS measurement of $Zh$ production, with $h\to b\bar b$ and $Z$ decaying into leptons, corresponding to an integrated luminosity of 139 fb$^{-1}$~\cite{ATLAS:2020fcp}.\footnote{The measurement of $pp\to Zh$ has also been performed by the CMS collaboration~\cite{CMS:2023vzh}, and more recently discussed by ATLAS in Ref.~\cite{ATLAS:2024yzu}. We focus on Ref.~\cite{ATLAS:2020fcp} because it provides detailed information about the number of events in each category.}  In the ATLAS analysis, events were categorized into 0-lepton  ($0\ell$) and 2-lepton ($2\ell$)  channels, with $2\ell$ being $e^+e^-$ or $\mu^+\mu^-$. All events were required to have exactly two $b$-tagged jets, and 0, 1, $\geq1$ of untagged jets were accepted in different categories. Up to three regions of the transverse momentum of the $Z$-boson, $p_T^V$, were considered, leading altogether to 10 distinct signal regions. We will focus on the 8 signal regions with higher $p_T^V$ listed in Tables~\ref{tab:0l_signal_regions_139} and \ref{tab:2l_signal_regions_139}, since the new physics contributions to the low $p_T^V$ bins are very small.

\begin{table*}[t]
\centering
\renewcommand{\arraystretch}{1.25}
\setlength{\tabcolsep}{2pt}
\begin{tabular}{lcccc}
\hline
Signal regions
& \multicolumn{2}{c}{0-lepton}
& \multicolumn{2}{c}{0-lepton} \\
& \multicolumn{2}{c}{$150~\mathrm{GeV} < p_T^V < 250~\mathrm{GeV}$}
& \multicolumn{2}{c}{$p_T^V > 250~\mathrm{GeV}$} \\
\hline
Sample
& 2-jet & 3-jet
& 2-jet & 3-jet \\

\hline 
SM $pp\to Zh$ 

& $144 \pm 24$ & $128 \pm 22$ & $39 \pm 6$ & $32 \pm 6$ \\

Other bkg. 

& $5234 \pm 63$ & $11910 \pm 100$ & $554 \pm 15$ & $1011 \pm 21$ \\

\hline 

Total bkg.

& $5378 \pm 67$ & $12038 \pm 102$ & $593 \pm 16$ & $1043 \pm 22$ \\

\hline
Data

& $5397$ & $11875$ & $578$ & $1046$ \\
\hline
\hline
Signal
&184 &164 &201 &187 \\
$Z$
&1.65 & 2.18 & 7.05  & 4.60 \\
\hline
\end{tabular}
\caption{The signal corresponding to the opposite-sign $g_R^{Zbb} = -g_{R,SM}^{Zbb}$ scenario and its statistical significance, background, and data for each signal region in the 0-lepton channel after the full selection. The SM $pp \to Zh$ for $\mu = 1.00$, other background (consisting of $Z$+jets, $W$+jets, single top, $t\bar t$ and diboson), and data yields are obtained from the ATLAS search for $pp \to Zh$ corresponding to an integrated luminosity of $139 \;{\rm fb^{-1}}$~\cite{ATLAS:2020fcp}, Table 10. The number of signal events is obtained from Eq.~(\ref{eq:N}) and the statistical significance is obtained by the procedure summarized in Appendix A. 
}
\label{tab:0l_signal_regions_139}
\end{table*}

\begin{table*}[t]
\centering
\renewcommand{\arraystretch}{1.25}
\begin{tabular}{lcccccc}
\hline
Signal regions
& \multicolumn{2}{c}{2-lepton}
& \multicolumn{2}{c}{2-lepton} \\
& \multicolumn{2}{c}{$150~\mathrm{GeV} < p_T^V < 250~\mathrm{GeV}$}
& \multicolumn{2}{c}{$p_T^V > 250~\mathrm{GeV}$} \\
\hline
Sample
& 2-jet & $\geq 3$-jet
& 2-jet & $\geq 3$-jet \\

\hline 

SM $pp\to Zh$ 

& $33 \pm 6$ & $58 \pm 12$ & $10 \pm 2$ & $18 \pm 3$ \\

Other bkg.

& $851 \pm 19$ & $3058 \pm 44$ & $106 \pm 4$ & $605 \pm 14$ \\

\hline 
Total bkg.

& $884 \pm 20$ & $3116 \pm 46$ & $116 \pm 5$ & $623 \pm 14$ \\

\hline
Data

& $881$ & $3148$ & $123$ & $614$ \\
\hline
\hline
Signal 
&35 & 55 & 42 &76 \\
$Z$
&1.06 &0.32 &2.70 & 2.89 \\
\hline
\end{tabular}
\caption{The same as Table~\ref{tab:0l_signal_regions_139} for the 2-lepton channel. The corresponding table in the ATLAS analysis is Table 11~\cite{ATLAS:2020fcp}. }
\label{tab:2l_signal_regions_139}
\end{table*}

After the event selection, the dominant background processes are $Z$+jets and $t \bar t$. Several variables describing the kinematics, jet flavor, and missing transverse momentum of signal events were constructed, and multivariate discriminants were used to maximize the sensitivity to the $Zh$ signal. After the full selection obtained by the boosted decision tree (BDT) analysis, the expectations for the $Zh$ signal in the SM, the background, together with observed data for each signal region were summarized in Tables 10 and 11 of Ref.~\cite{ATLAS:2020fcp}, and are also reproduced in Tables~\ref{tab:0l_signal_regions_139} and \ref{tab:2l_signal_regions_139}. Note that the process $pp\to Zh$ in the SM (which was the signal for the ATLAS analysis) becomes a part of the background for our study. 

The SMEFT interactions are implemented by extending the SM Lagrangian in a \textsc{FeynRules}~\cite{Alloul:2013bka} model file. The corresponding UFO model is then generated using \textsc{FeynRules} and \textsc{NLOCT}~\cite{Degrande:2014vpa}.
The signal and the SM contribution to $pp\to Zh$ are generated at next-to-leading order (NLO) in QCD using \textsc{MadGraph5\_aMC@NLO}~\cite{Alwall:2014hca}, including up to one additional jet with the 0 and 1 jet events merged. Jet merging is performed using the FxFx merging scheme~\cite{Frederix:2012ps} with $\textrm{Qcut} = 30\;{\rm GeV}$. The gluon fusion contribution to the $Zh$ final state for both signal and the SM is  generated separately at LO in QCD, rescaled to match the NLO cross section of $118.4\;{\rm fb}$~\cite{CampilloAveleira:2025rbh}, and then added to $Zh$ data.
Decays are generated by \textsc{MadSpin}~\cite{Artoisenet:2012st}. All events are produced in the five flavor (5F) scheme, in which the bottom quark is treated as a massless parton inside the proton. Parton showering and hadronization are simulated with \textsc{Pythia~8}~\cite{Sjostrand:2007gs}, while detector simulations are performed using \textsc{Delphes}~\cite{deFavereau:2013fsa}. 
Signal events are generated for various values of Wilson coefficients in Eqs.~(\ref{eq:del_gL}) and (\ref{eq:del_gR}), while the SM limit is recovered by setting all of them to zero. The SM setup is also used to validate the cross sections against \textsc{MadGraph5} benchmarks.

For all considered signal regions, we find excellent agreement between distributions of all kinematic variables entering the BDT analysis for the signal and the SM $pp\to Zh$ except for 
the invariant mass of two $b$-jets, $m_{bb}$, which besides the peak near the $m_h$ expected from the SM  $pp\to Zh$ also has a tail extending to larger values. This can be understood from the fact that a part of our signal has 3 hard $b$-jets (see Fig.~\ref{fig:smeft_diags} (c)) which will still be kept by the initial event selection if one of them is not tagged as a $b$-jet. Two thirds of such events do not have $b$-tagged jets originating from the Higgs boson, and so the resulting $m_{bb}$ is typically much larger than $m_h$ due to higher $p_T$ of the jets. To remove this part of the signal, we impose 
$m_{bb}<150\;{\rm GeV}$ in addition to all the basic selection criteria of the ATLAS analysis. Within that range, even the distribution of $m_{bb}$ agrees well with the corresponding distribution of the SM $pp\to Zh$. Based on this agreement, we assume equal BDT selection efficiencies for the SM $pp\to Zh$ and for our signal. From the generated number of events for $pp\to Zh$ assuming non-zero Wilson coefficients, $N(C)$, the generated number of events  in the SM, $N(0)$, and the corresponding number of events reported by ATLAS after the BDT analysis, $N(0)_{\textrm{BDT}}$, we obtain the number of expected signal events corresponding to modified $Zb\bar b$ couplings that would pass the BDT analysis as
\begin{equation}
N_{\textrm{signal}}(C)  = \left( \frac{N(0)_{\textrm{BDT}}}{N(0)} \right) N(C) - N(0)_{\textrm{BDT}}.
\label{eq:N}
\end{equation}
We repeat this for every considered signal region. 

The numbers of expected signal events corresponding to the opposite-sign $g_R^{Zbb} = -g_{R,SM}^{Zbb}$ for each signal region, together with their statistical significances, $Z$, are listed in Tables~\ref{tab:0l_signal_regions_139} and \ref{tab:2l_signal_regions_139}. Details of the statistical analysis are summarized in Appendix A. The significances are combined in quadrature across the signal regions, resulting in $Z=9.74$. The combined results across all signal regions therefore exclude the opposite-sign $g_R^{Zbb}$ scenario at a significance exceeding $9.5 \sigma$.

\begin{figure}[t]
    \includegraphics[width=3in]{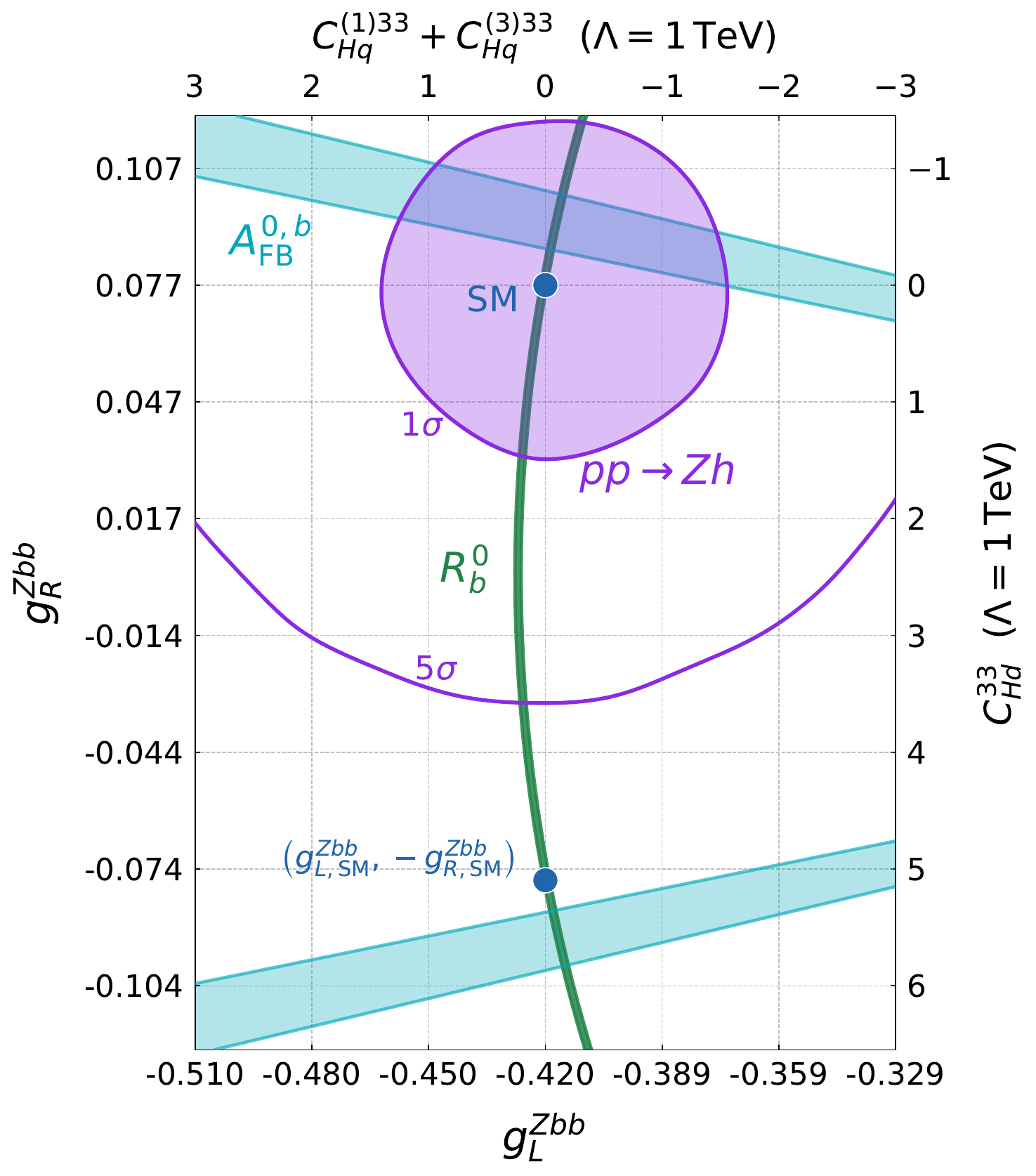}
    \caption{Contours of constant CLs, $1\sigma$ and $5\sigma$,  in the $(C_{Hq}^{(1)33}+C_{Hq}^{(3)33} ,\, C_{Hd}^{33})$ plane, or the $(g_L^{Z bb},\, g_R^{Z bb})$ plane, derived from the ATLAS search for $pp \to Zh$ corresponding to an integrated luminosity of $139 \;{\rm fb^{-1}}$~\cite{ATLAS:2020fcp}. We also show $1\sigma$ contours of $R_b^0$ and $A_{FB}^{0,b}$, see the text.
   \label{fig:CL_current}}
\end{figure}

We extend the analysis to different values of the Wilson coefficients corresponding to modifications of both $Zb\bar b$ couplings, and plot contours of constant confidence levels (CL), $1\sigma$ and $5\sigma$, in the $(C_{Hq}^{(1)33}+C_{Hq}^{(3)33} ,\, C_{Hd}^{33})$ or $(g_L^{Z bb},\, g_R^{Z bb})$ plane in Fig.~\ref{fig:CL_current}. The correspondence between Wilson coefficients and the couplings is given in Eqs.~(\ref{eq:gL}), (\ref{eq:gR}) and (\ref{eq:del_gL}), (\ref{eq:del_gR}), and assumes no contributions from operators of dimension-8 and higher. In addition, we assume that $\delta g_L^{Z tt} = 0$, which corresponds to $C_{Hq}^{(1)33}-C_{Hq}^{(3)33} = 0$, see Eq.(\ref{eq:Ztt}). This combination of Wilson coefficients is constrained from the $t\bar{t}Z$ production at the LHC, and the one-operator-at-a-time fit to data, allowing contributions up to $1/\Lambda^4$, finds $(C_{H q}^{(1) 33} - C_{Hq}^{(3) 33}) \in [-1.147,1.585]$~\cite{Ethier:2021bye}. Non-zero $\delta g_L^{Z tt}$ contributes to the signal only through the diagram in Fig.~\ref{fig:ggzh_diags} (c). Varying $(C_{H q}^{(1) 33} - C_{Hq}^{(3) 33})$ within the allowed range results in barely visible differences in the exclusion limits, and thus we do not indicate them.

In Fig.~\ref{fig:CL_current}, we also show $1\sigma$ contours of $Z$-pole observables obtained from LEP and SLC measurements: the ratio of the partial decay width of the $Z$ boson into a bottom-quark pair to its total hadronic decay width, $R_b^0$, and the bottom-quark forward-backward asymmetry, $A_{\mathrm{FB}}^{0,b}$.\footnote{The experimental values together with formulas converting them to the $Zb\bar b$ couplings can be found in Ref.~\cite{ParticleDataGroup:2026aaa}. The evaluation of $A_{\mathrm{FB}}^{0,b}$ includes finite bottom-quark mass corrections~\cite{FIELD_1998}. Similarly, the evaluation of $R_b^0$ includes QED and QCD radiative corrections, together with finite bottom-quark mass corrections through $\mathcal{O}(m_b^2/M_Z^2)$ and the axial singlet top-quark contribution at $\mathcal{O}(\alpha_s^2)$~\cite{Bernabeu:1990ws}. The running electromagnetic coupling includes leptonic vacuum-polarization corrections~\cite{Steinhauser_1998}, five-flavor hadronic contributions, and the conversion to the $\overline{\mathrm{MS}}$ scheme~\cite{ParticleDataGroup:2026aaa}.} The overlapping regions of couplings inferred from these two observables illustrate the discrepancy from the SM prediction and the existence of two solutions. This so-called $A_{\mathrm{FB}}^{0,b}$ anomaly, although not the subject of this paper, generated much attention; see for example, Refs.~\cite{Choudhury:2001hs,He:2002ha,Djouadi:2006rk,DaRold:2010as,Dermisek:2011xu,Dermisek:2012qx,Liu:2017xmc,Cheung:2020vqm,Crivellin:2020oup,Jurciukonis:2021wny,Fontes:2022sjp} with some models utilizing the opposite-sign solution, $g_R^{Zbb}  \lesssim -g_{R,SM}^{Zbb}$. Note that the two solutions, or even the $A_{\mathrm{FB}}^{0,b}$ anomaly, are often not reflected in SMEFT fits to data. Ref.~\cite{Breso-Pla:2021qoe} performed a global fit to the data,  but up to linear order in vertex corrections, which by construction cannot probe the opposite-sign solution. Other fits assume one operator at a time, in which case both $Zb\bar{b}$ couplings or the corresponding operators are mostly constrained by the much more precise $R_b^0$ measurement; see, for example, Ref.~\cite{CMS:2025ugn}.

Close to circular shapes of CLs centered on the SM in Fig.~\ref{fig:CL_current} confirm that the signal is proportional to $|\delta g_L^{Zbb}|^2 + |\delta g_R^{Zbb}|^2$ and thus dominated by the contact interactions squared as discussed in Sec.~\ref{sec:ModCouplings}. We find that the ATLAS analysis constrains the ranges of Wilson coefficients to 
\begin{eqnarray}
	(C_{Hq}^{(1)33}+C_{Hq}^{(3)33}) &\in [-1.56,1.40] \\
	C_{Hd}^{33}& \in [-1.40,1.48] \label{eq:ATLAS_CHd33}	
\end{eqnarray}
for $\Lambda = 1$~TeV at $1\sigma$. Under the assumption that only these dimension-6 operators contribute to the process and modifications of couplings, the results translate to $g_L^{Zbb} \in [-0.462,-0.372]$ and $ g_R^{Zbb} \in [0.032,0.119]$ at $1\sigma$.

Although these results are not competitive with LEP measurements and do not shed light on the $A_{FB}^{0,b}$ anomaly, they are obtained from the analysis that was not designed to be the most efficient for such a task. Most of the signal originating from Fig.~\ref{fig:smeft_diags} (c) does not contribute to this analysis. However, this analysis was able to rule out the opposite-sign $g_R^{Zbb}$ scenario that LEP measurements could not.

\section{Proposed $pp\to Zhb$ search} 

\label{sec:3b}

To take full advantage of the energy growth of the signal resulting from the contact interactions and the fact that most of the signal events have 3 $b$-jets (see Fig.~\ref{fig:smeft_diags} (c)), we also design a simple cut-based analysis focusing on the higher range of transverse momenta of final state particles. We consider two signal categories: 0-lepton with a veto on charged leptons, and exactly two same-flavor opposite-sign leptons. Both categories require at least 3 $b$-tagged jets, $N_b\geq 3$. Other selection criteria are listed in Table~\ref{tab:selection_140_3000} and consist of the following quantities: the transverse momenta of the leading and subleading leptons, $p_T^{\ell_{1,2}}$ ; the dilepton invariant mass and transverse momentum, $m_{\ell\ell}$ and $p_T^{\ell\ell}$, respectively; the invariant mass of the pair of $b$-tagged jets with the smallest $\Delta R$, $m_{bb}$;  the number of  non-$b$-tagged jets, $N_{\mathrm{non}\text{-}b}$; the magnitude of the missing transverse momentum,  $E_T^{\mathrm{miss}}$; the minimum of transverse masses formed from any selected $b$-tagged jet and the missing transverse momentum, $\min m_T(b,E_T^{\mathrm{miss}})$, where $m_T^2(b,E_T^{\mathrm{miss}})=2p_T^b E_T^{\mathrm{miss}}[1-\cos\Delta\phi(b,E_T^{\mathrm{miss}})]$~\cite{ATLAS:2017weo}; and the transverse mass  formed from the selected $bb$ system and the missing transverse momentum, $m_T(bb,E_T^{\mathrm{miss}})$, defined by $m_T^2(bb,E_T^{\mathrm{miss}})=m_{bb}^2+2[E_T^{bb}E_T^{\mathrm{miss}}-\mathbf{p}_T^{bb}\cdot\mathbf{p}_T^{\mathrm{miss}}]$, where $E_T^{bb}=\sqrt{m_{bb}^2+(p_T^{bb})^2}$~\cite{ATLAS:2022enb}.\footnote{The angular separation, $\Delta R$, is defined as $\Delta R=\sqrt{(\Delta\eta)^2+(\Delta\phi)^2}$ where $\eta$ and $\phi$ denote the pseudorapidity and the azimuthal angle, respectively. Central jets must satisfy $p_T>20\,\mathrm{GeV}$ and $|\eta|<2.5$, while forward jets satisfy $p_T>30\,\mathrm{GeV}$ and $2.5<|\eta|<4.5$. $N_b$  counts only the central $b$-tagged jets while $N_{\mathrm{non}\text{-}b}$ counts jets in both regions. Electrons and muons must satisfy $p_T>7\,\mathrm{GeV}$ with $|\eta|<2.47$ and $|\eta|<2.7$, respectively.} The last two variables are expected to efficiently remove $t\bar{t}$ and low $Zh$ invariant mass backgrounds, respectively. The selection criteria in Table~\ref{tab:selection_140_3000} efficiently suppress the background for an integrated luminosity of $140\,\mathrm{fb}^{-1}$  assumed in this section. The entries in parentheses are used for projections at HL-LHC discussed in Sec.~\ref{sec:HL-LHC}.

\begin{table}[htbp]
	\centering
	\setlength{\tabcolsep}{2pt}
	\renewcommand{\arraystretch}{1.15}
	\begin{tabular}{@{}lcc@{}}
		\hline
		\textbf{Selection}
		& \makebox[0.25\columnwidth][c]{\textbf{0-lepton }}
		& \makebox[0.25\columnwidth][c]{\textbf{2-lepton}} \\
		\hline
		$N_b$ & \multicolumn{2}{c}{$\geq 3$} \\
		$p_T^{\ell_{1,2}}$ & -- & $>60\,(65)\,\mathrm{GeV}$ \\
		$m_{\ell\ell}$ & -- & $81\leq m_{\ell\ell}\leq101\,\mathrm{GeV}$ \\
		$p_T^{\ell\ell}$ & -- & $>200\,(250)\,\mathrm{GeV}$ \\
				$m_{bb}$ & \multicolumn{2}{c}{$110\,(115)<m_{bb}<140\,(135)\,\mathrm{GeV}$} \\
		$N_{\mathrm{non}\text{-}b}$ & $\leq 2$ & $\leq 2\;(<2)$ \\
		$E_T^{\mathrm{miss}}$ & -- & -- $ \;(<200\,\mathrm{GeV})$ \\
		$\min m_T(b,E_T^{\mathrm{miss}})$ & $>150\,\mathrm{GeV}$ & -- \\
		$m_T(bb,E_T^{\mathrm{miss}})$ & $>500\,\mathrm{GeV}$ & -- \\
		\hline
	\end{tabular}
	\caption{Selection criteria for the 0-lepton and 2-lepton signal categories of the proposed $pp\to Zhb$ search assuming an integrated luminosity of $140\,\mathrm{fb}^{-1}$. If the selection criteria differ for HL-LHC, we show them in parentheses.	
	}
	\label{tab:selection_140_3000}
\end{table}

Similarly to the ATLAS analysis, we consider three main backgrounds,  $t\bar{t}$, $Z$+jets, and the SM contribution to $Zh$, for which we generate events at NLO in QCD. The  $t\bar{t}$ background is generated with up to two additional jets and merged using FxFx merging scheme. The $Z$+jets cross section is very large, $\sigma_{Z+\textrm{jets}}\sim 5.8\times 10^7\;{\rm fb}$, and to improve the event generation efficiency, we generate it differently for the 2-lepton and 0-lepton cases. For the 2-lepton case, we generate $\ell\ell$  with up to two additional jets in the 5F scheme and merge via FxFx merging scheme. We impose parton-level cuts $p_T^l > 50\;{\rm GeV}$ and $m_{ll} > 75$ GeV to improve the event generation efficiency. For the 0-lepton case, where the $p_T$ cut cannot be applied, we instead generate $pp\to Zb\bar b$ in the four flavor (4F) scheme, where the bottom quark is treated as massive and is not included in parton distribution functions. A previous study shows that the 4F and 5F schemes agree reasonably well in distributions of kinematic variables, but the total number of events in the 4F scheme is smaller compared to the data by about 20\%~\cite{Krauss:2016orf}. Thus, we rescale the generated number of events to match the data. Finally, the SM $Zh$ is generated in the same way as in Sec.~\ref{sec:ATLAS_CURRENT}.

\begin{figure}[t]
    \includegraphics[width=3in]{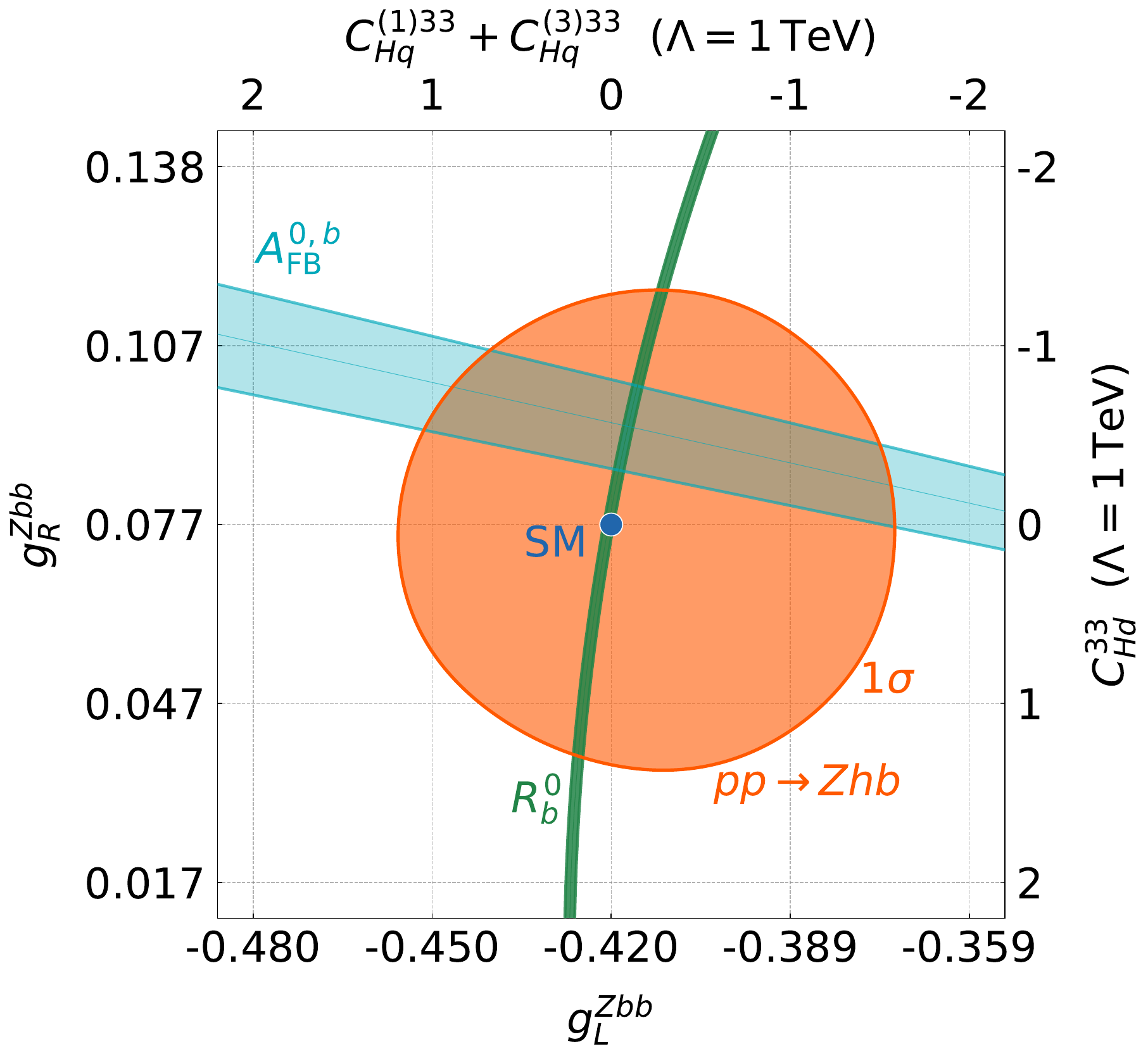}
    \caption{Contour of expected $1\sigma$ exclusion limits in the $(C_{Hq}^{(1)33}+C_{Hq}^{(3)33} ,\, C_{Hd}^{33})$ or $(g_L^{Z bb},\, g_R^{Z bb})$ plane,  derived from the cut-based $3b$-jets analysis of $pp \to Zhb$  corresponding to an integrated luminosity of $140~\mathrm{fb}^{-1}$. Also shown are $1\sigma$ contours of $R_b^0$ and $A_{FB}^{0,b}$.
    \label{fig:CL_3b_current}}
\end{figure}

The contour of expected $1\sigma$ CL corresponding to an integrated luminosity of $140 \;{\rm fb^{-1}}$  is plotted in Fig.~\ref{fig:CL_3b_current}. As in the previous section, we assume that $\delta g_L^{Z tt} = 0$, which corresponds to $C_{Hq}^{(1)33}-C_{Hq}^{(3)33} = 0$, but we have checked that within the allowed range, $(C_{H q}^{(1) 33} - C_{Hq}^{(3) 33}) \in [-1.147,1.585]$~\cite{Ethier:2021bye}, results do not visibly change. This was expected since the diagram in Fig.~\ref{fig:ggzh_diags} (c) contributes even less to the $3b$ final state.

From Fig.~\ref{fig:CL_3b_current} we see that this simple cut-based analysis, including only 2 signal regions, performs comparably to the sophisticated $pp\to Zh$ analysis in Sec.~\ref{sec:ATLAS_CURRENT}, and thus, with further improvements, including dedicated multivariate analysis and more signal regions, is expected to lead to better sensitivity. In the next section, we will see that with the larger data sets at HL-LHC it will become superior.

\section{Sensitivity to $Zb\bar b$ couplings at HL-LHC} 
\label{sec:HL-LHC}

\begin{figure}[t]
    \includegraphics[width=3in]{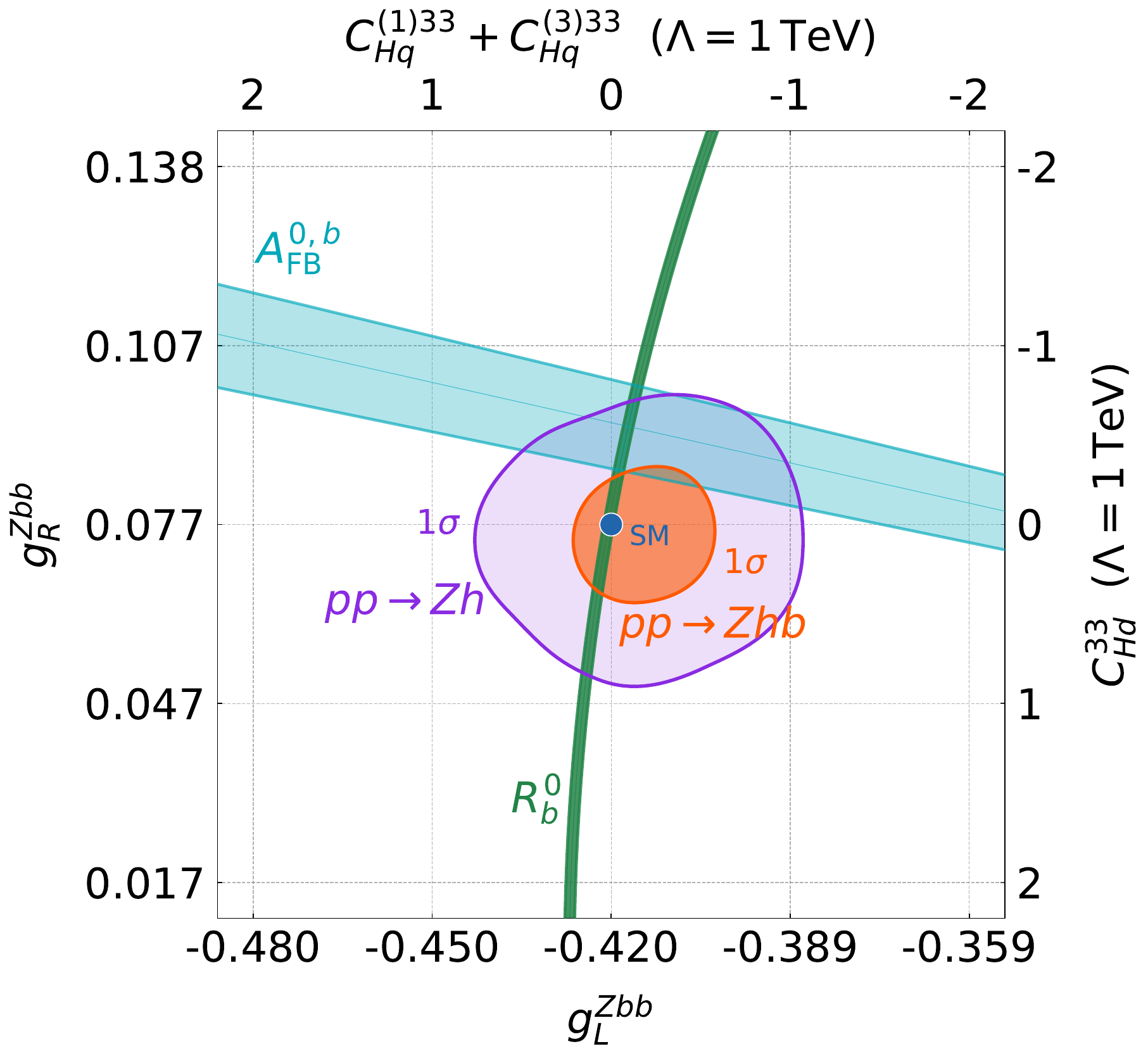}
  \caption{Contours of expected $1\sigma$ exclusion limits of the ATLAS $pp\to Zh$ analysis and the proposed $pp\to Zhb$ analysis for an integrated luminosity of 3 ab$^{-1}$  at the HL-LHC in the $(C_{Hq}^{(1)33}+C_{Hq}^{(3)33} ,\, C_{Hd}^{33})$ or $(g_L^{Z bb},\, g_R^{Z bb})$ plane. Also shown are $1\sigma$ contours of $R_b^0$ and $A_{FB}^{0,b}$. \label{fig:CL_HL}  }  
\end{figure}

Projected sensitivities to $Zb\bar b$ couplings of the same ATLAS analysis and the proposed $3b$ analysis for an integrated luminosity of 3 ab$^{-1}$  at the HL-LHC are plotted in Fig.~\ref{fig:CL_HL}. 
For the ATLAS $pp\to Zh$ contour of $1\sigma$, we do not assume any changes in the analysis itself, just a larger number of events scaling linearly with the luminosity, $\mathcal L$, and improvements in experimental uncertainties scaling as $1/\sqrt{\mathcal L}$. For the $pp\to Zhb$ analysis discussed in Sec.~\ref{sec:3b}, we impose a few tighter cuts allowed by a larger data set. They are indicated in parentheses in Table~\ref{tab:selection_140_3000}. As before, we assume $C_{Hq}^{(1)33}-C_{Hq}^{(3)33} = 0$, but we have checked that within the allowed range, $(C_{H q}^{(1) 33} - C_{Hq}^{(3) 33}) \in [-1.147,1.585]$~\cite{Ethier:2021bye}, the $pp\to Zh$ contour barely changes, while the change in the $pp\to Zhb$ contour is not visible.

We see that $1\sigma$ contours of both analyses deviate from circular shapes centered on the SM. As a result of smaller Wilson coefficients that could be probed at HL-LHC, the interference terms are no longer negligible compared to the contact interactions squared. 

The proposed $pp \to Zhb$ analysis is significantly more sensitive to modifications of $Zb\bar b$ couplings: We find that it could constrain the ranges of Wilson coefficients to $(C_{Hq}^{(1)33}+C_{Hq}^{(3)33}) \in [-0.58,0.21]$ and $ C_{Hd}^{33} \in [-0.32,0.44]$ at $1\sigma$. Under the assumption that only these dimension-6 operators contribute to the process, the results translate to $g_L^{Zbb} \in [-0.426,-0.402]$ and $ g_R^{Zbb} \in [0.064,0.087]$ at $1\sigma$. The obtained ranges exceed the reach of other proposals~\cite{Gauld:2019doc, Bishara:2023qhe}, and thus, with future data, this indirect method can provide the strongest constraints on $Zb\bar b$ couplings at the LHC. Even more interesting is the shape of the 2-dimensional exclusion limit. We see that this method can even probe the $A_{FB}^{0,b}$ anomaly remaining from LEP era. 

Furthermore, the suggested cut-based analysis could be significantly improved as already mentioned in Sec.~\ref{sec:3b}. In addition, future improvement in b-tagging efficiency can potentially further enhance the sensitivity.

\section{Discussion}
\label{sec:discussion}

Results presented in the previous two sections assume that only $C_{Hq}^{(1)33}$, $C_{Hq}^{(3)33}$, and $C_{Hd}^{33}$ contribute to the  $pp\to Zh(b)$ processes
and affect $Zb\bar b$ couplings. However, other dimension-6 operators could contribute to the $pp\to Zh(b)$ processes but not modify the $Zb\bar b$ couplings, and other operators at dimension-8 or higher could contribute to the process or the couplings. 

Among dimension-6 operators that could contribute to the $pp\to Zh(b)$ processes are the same-type operators, $C_{Hq}^{(1)}$, $C_{Hq}^{(3)}$, and $C_{Hd}^{}$, that modify the $Z$-couplings of lighter quarks, and ones that modify the $ZZh$ vertex. Constraints on the corresponding Wilson coefficients from $pp \to Zh$ and $Wh$ were obtained by ATLAS~\cite{ATLAS:2020fcp} and CMS~\cite{CMS:2024ksn}. In the CMS study, it was found that the sensitivities to all Wilson coefficients considered except for $C_{Hq}^{(3)11}$ are dominated by quadratic effects of Wilson coefficients.\footnote{The exception for $C_{Hq}^{(3)11}$ can be understood from the fact that it contributes to the $Z$ couplings of left-handed up and down quarks in opposite directions: Since these couplings have opposite signs, $g_{L, \rm SM,tree}^{Zuu} \simeq 0.35$ and $g_{L, \rm SM,tree}^{Zdd} \simeq -0.42$, the couplings are simultaneously increased or decreased in magnitude by $C_{Hq}^{(3)11}$, maximizing possible interference effects. In contrast, $C_{Hq}^{(1)11}$ shifts both up and down couplings in the same direction, and thus possible interference effects largely cancel.} 
This means that the contributions of new physics mostly originate from the contact interactions-squared, which is similar to our findings from the contributions of $C_{Hq}^{(1)33}$, $C_{Hq}^{(3)33}$ and $C_{Hd}^{33}$. Since the contributions of new physics affecting the couplings of light quarks and the $b$-quark to $pp\to Zh(b)$ do not interfere, it is expected that non-zero Wilson coefficients of all operators except for $C_{Hq}^{(3)11}$ will lead to stronger constraints on $C_{Hq}^{(1)33} + C_{Hq}^{(3)33}$ and $C_{Hd}^{33}$. We have confirmed this for $C_{Hd}^{11}$. Finally, we allow non-zero $C_{Hq}^{(3)11}$ within the bounds obtained from one-operator-at-a-time fit~\cite{CMS:2024ksn,CMS:2025ugn} and find that in the direction where it destructively interferes with the SM amplitude, the allowed range of $C_{Hd}^{33}$ broadens from Eq.~\eqref{eq:ATLAS_CHd33} to $|C_{Hd}^{(3)33}|\lesssim 2$. Thus, we find that the results obtained in previous sections are quite robust to the presence of other dimension-6 operators.

Somewhat large value of the Wilson coefficient required for the opposite-sign $g_R^{Zbb} = -g_{R,SM}^{Zbb}$ scenario warrants the question of whether dimension-8 or higher operators could significantly alter the results. In the Warsaw basis, 28 and 11 operators contain $b$-quark left- and right-handed chiral currents at the dimension-8 level~\cite{Murphy:2020rsh}. Out of 11 operators containing right-handed currents, 8 operators do not modify the $Zb\bar b$ couplings and contribute only to $pp\to Z_T h$. Since our signal is dominated by $Z_L h$, it cannot be canceled by contributions from those operators. Among the remaining three operators, $\mathcal{O}_{d^2 H^4 D}$ can modify $g_R^{Zbb} $ and generate a $Z_L$-$h$-$b_R$-$\bar b_R$ contact interaction just like the dimension-6 operator we considered, but with different combinatoric factors; and  $\mathcal{O}_{d^2 H^2 D^3}^{(1)}$ and $\mathcal{O}_{d^2 H^2 D^3}^{(2)}$ in Table 5 of Ref.~\cite{Murphy:2020rsh} can only generate the contact interaction $Z_L$-$h$-$b_R$-$\bar b_R$ [they cannot modify the $Zb\bar b$ couplings]. Thus, it is in principle possible that $g_R^{Zbb}$ coupling is modified or even its sign is flipped, but the $pp \to Zh$ signal is cancelled between dimension-6 and dimension-8 or higher operators. However, this would require a dimension-8 operator with a correspondingly large Wilson coefficient that would lead to other signals of new physics, for example, $p p \to ZZ W^+ W^-,\;W^+ W^- h h$, or $ZZZh$, which feature small SM backgrounds. 
In addition, the scale of new physics required for a dimension-8 operator to flip the sign of $g_R^{Zbb}$, inferred from unitarity and analyticity constraints~\cite{Bresciani:2026acy}, is quite low, $ \Lambda \simeq 900$ GeV, and thus the corresponding UV theory could be directly probed at the LHC.

\section{Conclusions}

Under the assumption that possible modifications of $Zb\bar{b}$ couplings from SM predictions originate from dimension-6 SMEFT operators, we were able to rule out the opposite-sign scenario for the right-handed coupling,  $g_R^{Zbb} = -g_{R,SM}^{Zbb}$. We have recast the ATLAS search for $pp\to Zh$ corresponding to an integrated luminosity of 139 fb$^{-1}$~\cite{ATLAS:2020fcp} to which the dimension-6 operators would contribute through the $Z$-$h$-$b$-$\bar b$ contact interactions, and found that the opposite-sign scenario is ruled out by more than 9.5$\sigma$. 

We have also designed a simple cut-based analysis for $pp\to Zhb$, focusing on the higher range of transverse momenta of final state particles and requiring at least 3 $b$-jets, which is more characteristic of the signal resulting from the $Z$-$h$-$b$-$\bar b$ contact interactions, and found that it performs comparably to the sophisticated $pp\to Zh$ ATLAS analysis~\cite{ATLAS:2020fcp}. Thus, with further improvements including BDT  and more signal regions, it is expected to lead to better sensitivity. 

We have also presented projected sensitivities to $Zb\bar b$ couplings or the corresponding Wilson coefficients of both ATLAS and the proposed $pp\to Zhb$ analyses for an integrated luminosity of 3 ab$^{-1}$  at the HL-LHC. We have found that the proposed $pp \to Zhb$ analysis is significantly more sensitive to modifications of $Zb\bar b$ couplings and can provide the strongest constraints on $Zb\bar b$ couplings at the LHC, exceeding the reach of other proposals~\cite{Gauld:2019doc, Bishara:2023qhe}. It can even probe the $A_{FB}^{0,b}$ anomaly remaining from the LEP era. Furthermore, the suggested cut-based analysis could be significantly improved.

We have argued and verified that the presence of other dimension-6 operators that do not modify $Zb\bar{b}$ couplings but could contribute to the $pp \to Zh $ process does not invalidate our results. In fact, the presence of all operators except for $C_{Hq}^{(3)11}$ makes the exclusion limits on $Zb\bar{b}$ couplings stronger. The $C_{Hq}^{(3)11}$ operator, through the interference with SM contributions,  only slightly weakens the obtained limits. However, if dimension-8 operators contribute to $Zb\bar{b}$ couplings or $pp \to Zh(b)$ comparably to dimension-6 operators, the unique relations between modification of the couplings and the $Z$-$h$-$b$-$\bar b$ contact interactions are lost. Such a scenario would require a lower scale of new physics, opening the possibility of directly probing it at the LHC, and it would also lead to numerous new signals resulting from dimension-8 operators that could be independently constrained.  

We encourage experimental collaborations to include signal regions with $3b$-jets in future $pp\to Zh$ searches. Although they will not improve the measurement of the $pp\to Zh$ cross section, they can be used to efficiently constrain Wilson coefficients of operators modifying the $Zb\bar{b}$ couplings. With future data, it might also be possible to explore other decay modes of the Higgs boson, besides the $b\bar b$, which could make the $pp\to Zhb$ search even more sensitive.

\acknowledgments
R.D. would like to thank the Galileo Galilei Institute (GGI) for Theoretical Physics, Firenze, Italy, and the Grant Simons Visiting Program at GGI for hospitality and support during the completion of this work. The work of K.H. was supported in part by the U.S. National Science Foundation under grant PHY2310149. The work of N.M. is supported in part by the U.S. Department of Energy under grant No. DEFG02-13ER41976/DESC0009913, and the DOE QuantISED program through the theory consortium “Intersections of QIS and Theoretical Particle Physics” at Fermilab (FNAL 20-17) under contract No. 89243024CSC000002. This research was supported in part by Lilly Endowment, Inc., through its support for the Indiana University Pervasive Technology Institute. The authors acknowledge the use of OpenAI Codex to assist with debugging and verifying the analysis scripts. The authors take full responsibility for the accuracy and integrity of the work.

\appendix 

\section{Details of the statistical analysis}
In this Appendix, we detail on how the exclusion limits are obtained for both the ATLAS and our proposed $pp\to Zhb$ analyses. 
\subsection{ATLAS $pp\to Zh$ analysis}
In order to quantify how strongly each benchmark coupling is excluded, we use profile likelihood ratio test statistic while taking into account the uncertainty of the  background~\cite{Cowan_2011}. This is built from the likelihood
\begin{equation} 
\begin{split}
L(\mu,b) := & \prod_i \frac{(\mu s_i + b_i)^{n_i}}{n_i!} e^{-(\mu s_i + b_i)} \\
\times & \frac{1}{\sqrt{2\pi\sigma_i^2}} \exp\left[ -\frac{(b_i-b_i^0)^2}{2\sigma_i^2} \right],
\end{split}
\end{equation}
where $n_i$ is the observed number of events, $b_i^0$ is the number of background events assuming it is the SM with uncertainty $\sigma_i$, and $s_i$ the expected number of signal events where $i$ labels the signal regions. The signal strength $\mu$ is used to rescale the signal such that $\mu=0$ is the background-only hypothesis and $\mu=1$ is the signal and background hypothesis for each benchmark coupling that we test.

To test the agreement between the data and a given $\mu$, we use the statistical variable $q_\mu$ as defined in Ref.~\cite{Cowan_2011},
\begin{equation}
q_\mu= -2 \log \frac{L(\mu, \hat{\hat{b}})}{L(\hat{\mu},\hat{b})}.
\end{equation}
For a given $\mu$, $\hat{\hat{b}}$ maximizes $L$ while $\hat\mu$ and $\hat b$ maximize $L$ globally. Since we are only interested in testing a given coupling, we only need $q_1$ which is obtained by setting $\mu=1$.

From $q_1$, we can obtain $\textrm{CL}_s\equiv \textrm{CL}_{s+b}/\textrm{CL}_{b}$ where $ \textrm{CL}_{s+b}$ and $ \textrm{CL}_{b}$ are confidence levels (CL) under signal-plus-background and background-only hypotheses. Due to large sample sizes of $\mathcal{O}(100)$ or larger, we evaluate the confidence level at $\mu=1$ with the asymptotic formula~\cite{Cowan_2011}
\begin{equation}
	\textrm{CL}_s := \frac{1-\Phi\left(\sqrt{q_1}\right)}{\Phi\left(\sqrt{q_1^A}-\sqrt{q_1}\right)},
\end{equation}
where
\begin{equation}
  \Phi(x) = \frac{1}{\sqrt{2\pi}}\int_{-\infty}^{x} e^{-t^2/2}\,dt,
  \label{eq:Phi}
\end{equation}
is the cumulative distribution function of the normal distribution. $q_1^A$ is obtained by replacing the observed number of events by the expected background yields, $n_i\to b_i^0$. For the current luminosity, $\mathcal{L}=139\;{\rm fb}^{-1}$, we take advantage of the number of events observed and calculate the observed exclusion significance via
\begin{equation}
Z_{\rm obs} = \sqrt{q_{1}}.
\end{equation}

The exclusion limit with a significance of $m$ standard deviations is obtained when
\begin{equation}
	\textrm{CL}_s < 2 [1-\Phi(m)].
	\label{eq:exclusion_limit}
\end{equation}
This translates to $\textrm{CL}_s<0.32$ and $\textrm{CL}_s<5.7\times10^{-7}$ for $m=1$ and $5$, respectively.

For projected limits at HL-LHC, $\mathcal{L}=3\;{\rm ab}^{-1}$, we use the expected exclusion significance,
\begin{equation}
Z_{\mathrm{exp}} = \sqrt{q_1\big|_{n_i=b_i^0}},
\end{equation}
where we assume the number of events observed is equal to SM prediction, $n_i\to b_i^0$. We first simulate the events at current luminosity and then rescale the events to obtain the expected yield at HL-LHC: $s_i \to R_{\mathcal{L}} s_i$ and $b_i^0 \to R_{\mathcal{L}} b_i^0$, and $\sigma_i \to \sqrt{R_{\mathcal{L}}}\,\sigma_i$. Here we have assumed the uncertainty scales as $1/\sqrt{\mathcal{L}}$ and $R_{\mathcal{L}}=3000/139$. 

The number of observed events and the estimated backgrounds are summarized in Tables \ref{tab:0l_signal_regions_139} and \ref{tab:2l_signal_regions_139} for $139\;{\rm fb^{-1}}$ and $3\;{\rm ab^{-1}}$.

\subsection{Proposed $pp\to Zhb$ search}
For our proposed analysis $pp\to Zhb$, the number of events are relatively small in which case the previously used asymptotic approximation breaks down~\cite{Bhattiprolu_2021}. Therefore, we evaluate the confidence level, $\textrm{CL}_s$, from the exact Poisson probabilities while treating the predicted yields as fixed following Ref.~\cite{Junk:1999kv}. For the expected exclusion limit, we consider every possible outcome unlike the ATLAS analysis where we simply replaced $n_i\to b_i^0$. This is possible since the two signal regions are expected to have small yields and no nuisance parameters, and we enumerate all the outcomes with non-negligible probability. In the ATLAS analysis due to the number of signal regions and large number of yields, it is impractical to do such an enumeration. However, the large yields ensure the asymptotic formula to be accurate. 

In this analysis we neglect the uncertainty of the background. This leads to a simpler form of the likelihood, 
\begin{equation}
L(\mu)=\prod_{n_i \in \bold{n}} P(n_i| \mu s_i+b_i^0),
\end{equation}
where $P(n_i|\nu_i) = \nu^ne^{-\nu}/n!$ is the Poisson probability. 
The likelihood ratio for a set of event counts $\mathbf{n}$ is given as
\begin{equation}
	X(\mathbf{n})=\prod_{n_i \in \bold{n}} \frac{P(n_i| s_i+b_i^0)}{P(n_i\mid b_i^0)},
\end{equation}
The confidence levels are calculated from the probabilities of a likelihood ratio that are no larger than $X$,
\begin{eqnarray}
	\textrm{CL}_{s+b}(X)&=&\sum_{X(\mathbf{n}')\leq X}\prod_{n_i' \in \bold{n'}} P(n_i'| s_i+b_i^0), \\ 
	\textrm{CL}_{b}(X)&=&\sum_{X(\mathbf{n}')\leq X}\prod_{n_i' \in \bold{n'}} P(n_i'| b_i^0).
\end{eqnarray}
The expected exclusion limit is then obtained by taking the average over background-only outcomes,
\begin{equation}
	\langle \textrm{CL}_s\rangle = \sum_\mathbf{n} \prod_{n_i \in \bold{n}} P(n_i|b_i^0) \textrm{CL}_s[X(\mathbf{n})],
\end{equation}
which is used in Eq.~\eqref{eq:exclusion_limit} to obtain the exclusion signficance. The infinite Poisson sums are truncated when the retained probability exceeds $1-10^{-10}$ under both hypotheses.

\bibliography{Zbb-references} 
\end{document}